\documentclass[%
 reprint,
superscriptaddress,
 amsmath,amssymb,
aps,
pra,
]{revtex4-2}

\usepackage{graphicx}
\usepackage{float}
\usepackage{subfig}
\newcommand{\mysize}{7.5cm}
\usepackage{graphicx}
\usepackage{dcolumn}
\usepackage{bm}
\usepackage{hyperref}

\begin{document}

\preprint{APS/123-QED}

\title{Four-state reference-frame-independent quantum key distribution with non-qubit sources}%

\author{Zhenhua Li}
\affiliation{School of Science and State Key Laboratory of Information Photonics and Optical Communications, Beijing University of Posts and Telecommunications, Beijing 100876, China}

\author{Tianqi Dou}
\affiliation{School of Science and State Key Laboratory of Information Photonics and Optical Communications, Beijing University of Posts and Telecommunications, Beijing 100876, China}

\author{Jipeng Wang}
\affiliation{School of Science and State Key Laboratory of Information Photonics and Optical Communications, Beijing University of Posts and Telecommunications, Beijing 100876, China}

\author{Zhongqi Sun}
\affiliation{School of Science and State Key Laboratory of Information Photonics and Optical Communications, Beijing University of Posts and Telecommunications, Beijing 100876, China}

\author{Fen Zhou}
\affiliation{School of Science and State Key Laboratory of Information Photonics and Optical Communications, Beijing University of Posts and Telecommunications, Beijing 100876, China}

\author{Yanxin Han}
\affiliation{School of Science and State Key Laboratory of Information Photonics and Optical Communications, Beijing University of Posts and Telecommunications, Beijing 100876, China}

\author{Yuqing Huang}
\affiliation{School of Science and State Key Laboratory of Information Photonics and Optical Communications, Beijing University of Posts and Telecommunications, Beijing 100876, China}

\author{Hongwei Liu}
\email{liuhw@itsec.gov.cn}
\affiliation{China Information Technology Security Evaluation Center, Beijing 100085, China}

\author{Haiqiang Ma}
\email{hqma@bupt.edu.cn}
\affiliation{School of Science and State Key Laboratory of Information Photonics and Optical Communications, Beijing University of Posts and Telecommunications, Beijing 100876, China}

\date{\today}

\begin{abstract}
  The discrepancy between theory and experiment severely limits the development of quantum key distribution (QKD). Reference-frame-independent (RFI) protocol has been proposed to avoid alignment of the reference frame. However, multiple optical modes caused by Trojan horse attacks and equipment loopholes inevitably result in the imperfect  non-qubit emitted signal. In this paper, we analyzed the security of the RFI-QKD protocol with non-qubit sources based on a generalizing loss-tolerant technique. The simulation results show that our work can effectively defend against non-qubit sources and other state preparation flaws. Moreover, it only requires the preparation of four quantum states, which reduces the complexity of the experiment.
\end{abstract}

\maketitle

\section{Introduction}

Quantum key distribution (QKD), based on the basic principles of quantum mechanics, ensures that the information interaction between the two communication parties (Alice and Bob) is information theoretic security \cite{PhysRevLett.67.661,lo1999unconditional,shor2000simple}. Since the first QKD protocol, BB84 protocol \cite{BB84}, was proposed, QKD has flourished not only in laboratories \cite{RN17,Muller_1993,doi:10.1080/09500349414552271,6471162} but also in companies \cite{company}. However, the gap between theory and experiment hinders the development of QKD, which is the reason why an untrusted third party Eve can eavesdrop secret keys. Considering that some of Eve's attacks are aimed at the detection side of the QKD system, such as detector time-shift attacks \cite{Qi2007TimeshiftAI,PhysRevA.78.042333}, faked states attacks \cite{PhysRevA.74.022313,Makarov2008FakedSA}, detector blinding attacks \cite{RN5,RN4}, etc., the proposal of measurement-device-independent QKD (MDI-QKD)  protocol \cite{PhysRevLett.108.130503} effectively eliminates the side channel information leakage at the detector side. The MDI-QKD protocol can be resistant to all attacks against detectors and be implemented experimentally \cite{Wang:17} using existing techniques.

The security loophole of the quantum source is more complex than the detector side. In practice, a weak coherent state (WCS) source is usually used as a substitute for a single photon source. By exploiting the multi-photon security vulnerability of WCS, Eve can perform photon number splitting (PNS) attacks \cite{PhysRevLett.85.1330}. As a countermeasure, the decoy state \cite{PhysRevLett.91.057901,PhysRevLett.94.230503,PhysRevLett.94.230504} scheme is proposed and can substantially improve the performance of QKD. In most QKD systems, reference frames from both sides of the communication are required to be aligned. However, extra reference frame alignment operations not only increase the complexity and cost of the system but also degrade the performance of the in practical system. Fortunately, the reference-frame-independent (RFI) protocol \cite{PhysRevA.82.012304} is proposed to provide an effective solution to this problem and is applied directly to free-space\cite{PhysRevA.86.032322,Wabnig_2013}, chip-to-chip\cite{PhysRevLett.112.130501}, and MDI protocol systems\cite{8026131}. Furthermore, state preparation flaws (SPF) caused by Alice's imperfect encoding system are always inevitable. To reduce the influence of the SPF, the Gottesman-Lo-Lütkenhaus-Preskill (GLLP) method \cite{1365172} can resist but it has a deficiency, which is the significant reduction in transmission distance or inability to generate secure keys.

Recently, the RFI-QKD with the loss-tolerant (LT) technique \cite{PhysRevA.90.052314,PhysRevA.92.042319} has been proposed to get rid of the deficiency of GLLP method. Moreover, the experimental results \cite{PhysRevA.99.032309,Liu:18} demonstrate that this protocol can resist both the SPF and a misaligned reference frame. However, in this protocol, there is also an unrealistic qubit assumption that the single photon signal sent by Alice is a qubit (qubit assumption), i.e., the encoding state must be in two-dimensional Hilbert space. This assumption is invalid when Eve launches Trojan horse attacks (THA) \cite{PhysRevA.73.022320,doi:10.1080/09500340108240904,PhysRevX.5.031030} on Alice. In a THA, Eve sends a bright light pulse containing Trojan photons to Alice's encoding system. The reflected back Trojan photons contain encoding information and they are sent to Eve, thus compromising the security of the system. Besides, an optical mode of the light pulse and Alice's coded information may be correlated due to the imperfection of Alice devices. In these cases, the single photon pulse sent by Alice is not a qubit (non-qubit assumption). As a consequence, Alice's encoding information may also be encoded to other dimensional spaces, and that is the high-dimensional information leakage. 

In this article, we investigate a method to consider the non-qubit sources of RFI-QKD protocol with the generalizing loss-tolerant (GLT) technique \cite{RN6}. It effectively avoids the SPF in a single-mode qubit subspace and information leakage in high-dimensional. In other words, it is not necessary to assume that the single photon pulse emitted by Alice is a qubit. Also, only four quantum states need to be prepared, i.e., two eigenstates of $Z$ basis and one of the eigenstates each in $X$ and $Y$ basis. Finally, we show the simulation of this method in the finite-key analysis.

\section{Method and security}
\subsection{RFI-QKD protocol}
In the case of unaligned reference frames, secret keys can be generated by using the RFI-QKD protocol \cite{PhysRevA.82.012304}. In the RFI-QKD protocol, only the $Z$ basis is assumed to be aligned, but the $X$ and $Y$ basis are instable. Alice and Bob can distill the secure keys in the $Z$ basis and estimate the maximum amount of information eavesdropped by Eve in $X$ and $Y$ basis. The $Z$, $X$ and $Y$ basis can be written as follows:
\begin{equation}
    \begin{aligned}
        Z_{A}&=Z_{B},\\
        X_{B}&=cos\omega X_{A} + sin \omega Y_{A},\\
        Y_{B}&=cos\omega Y_{A} - sin \omega X_{A},
    \end{aligned}
\end{equation}
where $\omega$ is the deviation of the reference frame which is unknown to Alice and Bob. The maximum amount of information $I_{E}^{U}$ eavesdropped by Eve can be bounded by 
\begin{equation}
    \begin{aligned}
        &I_E^U = \left( {1 - E_{ZZ}^{U}} \right)h\left( {\frac{{1 + {\upsilon _{\max }}}}{2}} \right) + E_{ZZ}^{U}h\left( {\frac{{1 + f\left( {{\upsilon _{\max }}} \right)}}{2}} \right)\\
&f\left( {{\upsilon _{\max }}} \right) = \frac{{\sqrt {\frac{{{C^L}}}{2} - {{\left( {1 - E_{ZZ}^{U}} \right)}^2}\upsilon _{\max }^2} }}{{E_{ZZ}^{U}}},\\
&{\upsilon _{\max }} = \min \left( {\frac{1}{{1 - E_{ZZ}^{U}}}\sqrt {\frac{{{C^L}}}{2}} ,1} \right),
    \end{aligned}
\end{equation}
when the upper bound of the bit error rate $E_{ZZ}^{U}$ in $Z$ basis is less than 15.9\%. It is obvious that $I_{E}$ is closely related to parameter $C$ and $E_{ZZ}^{U}$. The lower bound of $C$ can be expressed as follows:
\begin{equation}\label{CC}
    \begin{aligned}
    C&= {\left\langle {{X_A}{X_B}} \right\rangle ^2} + {\left\langle {{X_A}{Y_B}} \right\rangle ^2} + {\left\langle {{Y_A}{X_B}} \right\rangle ^2} + {\left\langle {{Y_A}{Y_B}} \right\rangle ^2} \\
     &= {\sum\limits_{\alpha ,\beta  \in \left\{ {X,Y} \right\}} {\left( {1 - 2E_{\alpha \beta }^{phase}} \right)} ^2}
    \end{aligned}
    \end{equation}
where $E^{phase}_{\alpha,\beta}$ characterizes the phase error rate when Alice and Bob choose $\alpha$ and $\beta$ basis respectively. Clearly, it is known from Eq. (\ref{CC}) that $C$ and $\beta$ are independent.

\subsection{Non-qubit sources}
The time-bin phase coding is assumed to be used in the following analysis. That is, time-bin encoding is applied to the $Z$ basis and phase encoding is applied to the $X$ and $Y$ basis. 

\emph{1.State preparation flaws.} In a single mode qubit space, we define the perfect quantum states of the $Z$ basis as $\left| {{0_Z}} \right\rangle$ and $\left| {{1_Z}} \right\rangle$, which correspond to the short and long path in the time-bin encoding, respectively. The four quantum states affected by the SPF can be expressed as:
\begin{gather}
\begin{split}
    \left| {{\phi _{0Z}}} \right\rangle  =& \cos \left( {\frac{{{\delta _{IM1}}}}{2}} \right)\left| {{0_Z}} \right\rangle  + \sin \left( {\frac{{{\delta _{IM1}}}}{2}} \right)\left| {{1_Z}} \right\rangle ,\\
    \left| {{\phi _{1Z}}} \right\rangle  =& \sin \left( {\frac{{{\delta _{IM2}}}}{2}} \right)\left| {{0_Z}} \right\rangle  + \cos \left( {\frac{{{\delta _{IM2}}}}{2}} \right)\left| {{1_Z}} \right\rangle ,\\
    \left| {{\phi _{0X}}} \right\rangle  =& \sin \left( {\frac{\pi }{4} + \frac{{{\delta _{BS1}}}}{2}} \right)\left| {{0_Z}} \right\rangle  \\
    &+ \cos \left( {\frac{\pi }{4} + \frac{{{\delta _{BS1}}}}{2}} \right){e^{i\left( {{\delta _{PM1}}} \right)}}\left| {{1_Z}} \right\rangle ,\\
    \left| {{\phi _{0Y}}} \right\rangle  =& \sin \left( {\frac{\pi }{4} + \frac{{{\delta _{BS2}}}}{2}} \right)\left| {{0_Z}} \right\rangle  \\
    &+ \cos \left( {\frac{\pi }{4} + \frac{{{\delta _{BS2}}}}{2}} \right){e^{i\left( {\frac{\pi }{2} + {\delta _{PM2}}} \right)}}\left| {{1_Z}} \right\rangle ,
    \end{split}
\end{gather}
where ${{\delta _{IM}}}$, ${{\delta _{BS}}}$ and ${{\delta _{PM}}}$ denote the SPF from the intensity modulator (IM), beam splitter (BS) and phase modulator (PM), respectively. Here, we assume that Alice prepares the four quantum states with equal probability. In other words, each quantum state is sent with a probability of 25\%. Their density matrices can be expanded by the Bloch vector ($P^{j\alpha }$), the identity operator (${{\hat \sigma }_I}$) and three Pauli operators (${{\hat \sigma }_X},{{\hat \sigma }_Y},{{\hat \sigma }_Z}$) as follows:
\begin{equation}
     \left| {{\phi _{j\alpha }}} \right\rangle \left\langle {{\phi _{j\alpha }}} \right| = \frac{1}{2}\left( {P_I^{j\alpha }{{\hat \sigma }_I}   + P_X^{j\alpha }{{\hat \sigma }_X} + P_Y^{j\alpha }{{\hat \sigma }_Y} + P_Z^{j\alpha }{{\hat \sigma }_Z}} \right).
\end{equation}

\emph{2. Non-qubit assumption.} Since there are multiple optical modes in the emitted signal due to the equipment imperfection, taking the polarization mode as an example, the formalism of the quantum state in the pulse sent to Bob can be written as follows:
\begin{equation}\label{polarization}
    {\left| {{\Omega _{j\alpha }}} \right\rangle _{B}} = \cos {\theta _{j\alpha }}{\left| {{\phi _{j\alpha }}} \right\rangle _{HB}} + \sin {\theta _{j\alpha }}{\left| {{\phi _{j\alpha }}} \right\rangle _{VB}}
\end{equation}
where $j$ and $\alpha$ are the bit value and the basis chosen randomly from Alice, respectively, with $j \in \left\{ {0,1} \right\}$ and $\alpha  \in \left\{ {X,Y,Z} \right\}$. The subscripts $H$ and $V$ are the horizontal and vertical polarization mode, respectively. The angle $\theta _{j\alpha }$ corresponds to the non-qubit assumption, which may be related to the choice of $j\alpha$. Moreover, we assume that state ${\left| {{\phi _{j\alpha }}} \right\rangle _{HB}} $ is a pure state in a single mode qubit space, and that state ${\left| {{\phi _{j\alpha }}} \right\rangle _{VB}}$ is any state outside of the single mode qubit space. Apparently, they are orthogonal to each other and their inner product is 0.

\emph{3. Trojan horse attacks.} Eve emits bright light into Alice's encoding system  and the reflected back Trojan photons are in the form of
\begin{equation}
    {\left| {{\xi _{j\alpha }}} \right\rangle _E} = {T_I}{\left| e \right\rangle _E} + {T_D}{\left| {{e_{j\alpha }}} \right\rangle _{E}},
\end{equation}
where ${\left| {{T_I}} \right|^2} + {\left| {{T_D}} \right|^2} = 1$. The quantum state ${\left| e \right\rangle _E}$ (${\left| {{e_{j\alpha }}} \right\rangle _E}$) means that it is unrelated (related) to the value $j\alpha$. Therefore, Eve can obtain the basis and bit information from the quantum state ${\left| {{e_{j\alpha }}} \right\rangle _E}$. Furthermore, in a coherent state, $ {\left| e \right\rangle _E}$ refers to a vacuum state (${\left| v \right\rangle _E}$), ${T_I}{\rm{ = }}{e^{ - {\raise0.5ex\hbox{$\scriptstyle \gamma $}
\kern-0.1em/\kern-0.15em
\lower0.25ex\hbox{$\scriptstyle 2$}}}}$ and ${T_D} = \sqrt {1 - {e^{ - \gamma }}}$, where $\gamma$ is the intensity of the reflected back Trojan photon. Considering the worst-case scenario, we assume that ${\left\langle {e}
\mathrel{\left | {\vphantom {e {{e_{j\alpha }}}}}
\right. \kern-\nulldelimiterspace}
{{{e_{j\alpha }}}} \right\rangle _E} = 0$, which means that ${\left| e \right\rangle _E}$ is orthogonal to ${\left| {{e_{j\alpha }}} \right\rangle _{E}}$.

After the above analysis, the quantum states with the SPF and non-qubit sources sent by Alice in each single photon pulse can be quantized as:
\begin{equation}\label{actual}
    \begin{aligned}
        &{\left| {{\Phi _{j\alpha }}} \right\rangle _{BE}} = {\left| {{\Omega _{j\alpha }}} \right\rangle _B} \otimes {\left| {{\xi _{j\alpha }}} \right\rangle _E}\\
        &= \cos {\theta _{j\alpha }}{T_1}{\left| {{\phi _{j\alpha }}} \right\rangle _{HB}}{\left| v \right\rangle _E} + \cos {\theta _{j\alpha }}{T_D}{\left| {{\phi _{j\alpha }}} \right\rangle _{HB}}{\left| {{e_{j\alpha }}} \right\rangle _E} \\&+ \sin {\theta _{j\alpha }}{\left| {{\phi _{j\alpha }}} \right\rangle _{VB}} \otimes \left( {{T_1}{{\left| v \right\rangle }_E} + {T_D}{{\left| {{e_{j\alpha }}} \right\rangle }_E}} \right),
    \end{aligned}
\end{equation}
where the first term (defined as $\left| \Gamma _{j\alpha}  \right\rangle $) is robust with the THA, but the others (defined as $\left| {{\Gamma_{j\alpha} ^ {\bot} }} \right\rangle $) are sensitive. 

\subsection{Calculation of parameter C}\label{Calculation of parameter C}
Because of the SPF and non-qubit sources, it is inappropriate that the phase error rate in $C$ is equal to the bit error rate. In the GLT process, the phase error rate is derived from the virtual process. In the following, we mainly show how to calculate $E_{XX}^{phase}$, since the phase error rate of other basis is calculated in a similar way. In the virtual process, Alice first prepares an entanglement state in the $Z$ basis:
\begin{equation}
    {\left| {{\Psi _Z}} \right\rangle _{ABE}} = \frac{1}{{\sqrt 2 }}\left[ {\left| 0 \right\rangle  \otimes {{\left| {{\Phi _{0Z}}} \right\rangle }_{BE}} + \left| 1 \right\rangle  \otimes {{\left| {{\Phi _{1Z}}} \right\rangle }_{BE}}} \right].
\end{equation}
 Alice and Bob measure it in $X$ basis successively. The bit error rate in the virtual state is the phase error rate, that is
\begin{equation}\label{Ephase}
    E_{XX}^{phase} = \frac{{Y_{1X,0X}^{Z,vir} + Y_{0X,1X}^{Z,vir}}}{{Y_{0X,0X}^{Z,vir} + Y_{1X,0X}^{Z,vir} + Y_{0X,1X}^{Z,vir} + Y_{1X,1X}^{Z,vir}}}.
\end{equation}
Here, $Y_{sX,jX}^{Z,vir}$ denotes the yield where Alice and Bob obtains $\left| {jX} \right\rangle $ and $\left| {sX} \right\rangle $  respectively, while Alice prepares ${\left| {{\Psi _Z}} \right\rangle _{ABE}}$ in the virtual process. Similarly, $s$ is the bit value with $s \in \left\{ {0,1} \right\}$. 

By measuring ${\left| {{\Psi _Z}} \right\rangle _{ABE}}$ with the $X$ basis, the state sent to Bob is
\begin{equation}\label{virBob}
    {\left| \psi  \right\rangle _{BE,jX}^{vir}}{\rm{ = }}\frac{1}{2}\left[ {\left| {{\Gamma _{0Z}}} \right\rangle  + \left| {\Gamma _{0Z}^ \bot } \right\rangle } \right.\left. { + {{( - 1)}^j}\left( {\left| {{\Gamma _{1Z}}} \right\rangle  + \left| {\Gamma _{1Z}^ \bot } \right\rangle } \right)} \right].
\end{equation}

\begin{widetext}
  In the form of Eq. (\ref{polarization}), Eq. (\ref{virBob}) can be normalized to follows:
  \begin{equation}
      \begin{aligned}
        \left| \psi  \right\rangle _{BE,jX}^{vir} = \frac{1}{2}\left( {\left| {\left| {{\Gamma _{0Z}}} \right\rangle {\rm{ + }}{{\left( { - 1} \right)}^j}\left| {{\Gamma _{1Z}}} \right\rangle } \right|{{\left| {{\gamma _{jX}}} \right\rangle }_{BE}} + \left| {\left| {\Gamma _{0Z}^ \bot } \right\rangle  + {{\left( { - 1} \right)}^j}\left| {\Gamma _{1Z}^ \bot } \right\rangle } \right|{{\left| {\gamma _{jX}^ \bot } \right\rangle }_{BE}}} \right)
      \end{aligned}.
  \end{equation}
  State ${\left| {{\gamma _{jX}}} \right\rangle _{BE}}$  corresponds to a single-mode qubit, and ${\left| {\gamma _{jX}^ \bot } \right\rangle _{BE}}$ corresponds to a state orthogonal to ${\left| {{\gamma _{jX}}} \right\rangle _{BE}}$ in any other mode. They are in the form of
  \begin{equation}
    \begin{aligned}
      {\left| {{\gamma _{jX}}} \right\rangle _{BE}}=\frac{{\left| {{\Gamma _{0Z}}} \right\rangle {\rm{ + }}{{\left( { - 1} \right)}^j}\left| {{\Gamma _{1Z}}} \right\rangle }}{{\left| {\left| {{\Gamma _{0Z}}} \right\rangle {\rm{ + }}{{\left( { - 1} \right)}^j}\left| {{\Gamma _{1Z}}} \right\rangle } \right|}},
      {\left| {\gamma _{jX}^ \bot } \right\rangle _{BE}} = \frac{{\left| {\Gamma _{0Z}^ \bot } \right\rangle  + {{\left( { - 1} \right)}^j}\left| {\Gamma _{1Z}^ \bot } \right\rangle }}{{\left| {\left| {\Gamma _{0Z}^ \bot } \right\rangle  + {{\left( { - 1} \right)}^j}\left| {\Gamma _{1Z}^ \bot } \right\rangle } \right|}}.
    \end{aligned}
  \end{equation}
  The yield of the virtual states can be expressed as
  \begin{gather}
      \label{yield_vir}
   \begin{split}
      &Y_{sX,jX}^{Z,vir} = \frac{1}{4}{P_{jX}}{\rm{Tr}}\left( {{{\hat D}_{sX}}\left| \psi  \right\rangle \langle \psi |_{BE,jX}^{vir}} \right)\\
       &= \frac{1}{4}{P_{jX}}\left[ {{\rm{Tr}}\left( {{F_j}{{\hat D}_{sX}}\left| {{\gamma _{jX}}} \right\rangle \langle {\gamma _{jX}}{|_{BE}}} \right) + {\rm{Tr}}\left( {{G_j}{{\hat D}_{sX}}\left| {{\gamma _{jX}}} \right\rangle \langle \gamma _{jX}^ \bot {|_{BE}} + G_j^*{{\hat D}_{sX}}\left| {\gamma _{jX}^ \bot } \right\rangle \langle {\gamma _{jX}}{|_{BE}} + {H_j}{{\hat D}_{sX}}\left| {\gamma _{jX}^ \bot } \right\rangle \langle \gamma _{jX}^ \bot {|_{BE}}} \right)} \right],
  \end{split}
  \end{gather}
  where ${P_{jX }}$ is the probability that Alice prepares an entanglement state $ {\left| {{\Psi _Z}} \right\rangle _{ABE}}$ and measures it to obtain
  the state $\left| {jX } \right\rangle $, ${\textstyle{1 \over 4}}$ is the probability Bob chooses $X$ basis and ${\hat D_{{s_X}}}$  is an operator that contains Bob's positive-operator valued measures (POVM) and Eve's action. Other coefficient expressions are detailed in the Appendix \ref{APPENDIX}. The first term can be expressed by a linear combination as follows:
  \begin{equation}
      \begin{aligned}
          {\rm{Tr}}\left( {{F_j}{{\hat D}_{sX}}\left| {{\gamma _{jX}}} \right\rangle {{\left\langle {{\gamma _{jX}}} \right|}_{BE}}} \right)= {F_j}\left( {P_I^{jX,vir}{q_{sX|Id}} + P_X^{jX,vir}{q_{sX|X}} + P_Y^{jX,vir}{q_{sX|Y}} + P_Z^{jX,vir}{q_{sX|Z}}} \right),
      \end{aligned}
  \end{equation}
  where ${q_{sX|t}} = {{{\rm{Tr}}\left( {{{\hat D}_{sX}}{\sigma _t}} \right)} \mathord{\left/
  {\vphantom {{{\rm{Tr}}\left( {{{\hat D}_{sX}}{\sigma _t}} \right)} 2}} \right.
  \kern-\nulldelimiterspace} 2}$, with $t \in \left\{ {I,X,Y,Z} \right\}$. The other items are calculated in a different way. The other terms can be re-represented as 
  \begin{equation}\label{nonqubit}
      \begin{aligned}
        {\rm{Tr}}\left( {{G_j}{{\hat D}_{sX}}\left| {{\gamma _{jX}}} \right\rangle \langle \gamma _{jX}^ \bot {|_{BE}} + G_j^*{{\hat D}_{sX}}\left| {\gamma _{jX}^ \bot } \right\rangle \langle {\gamma _{jX}}{|_{BE}} + {H_j}{{\hat D}_{sX}}\left| {\gamma _{jX}^ \bot } \right\rangle \langle \gamma _{jX}^ \bot {|_{BE}}} \right){\rm{ = Tr}}\left( {{{\hat D}_{sX}}{K_j}} \right),
      \end{aligned}
  \end{equation}
  where ${K_j}$ is matrix $\left[ {\begin{array}{*{20}{c}}
      {{H_j}}&{{G_{j}^{*}}}\\
      {{G_j}}&0
      \end{array}} \right]$,with two eigenvalues ${\lambda^{vir}_{\max,j}}$ and ${\lambda^{vir}_{\min,j}}$. In addition, the eigenvalue of ${{\hat D}_{sX}}$ is between 0 and 1 in the POVM. Therefore, $Y_{sX,jX}^{Z,vir}$ can be bounded by
  \begin{equation}\label{virstate}
      \begin{aligned}
          &\frac{1}{4}{P_{j\alpha }}\left[ {{F_j}\left( {P_I^{jX,vir}{q_{sX|Id}} + P_X^{jX,vir}{q_{sX|X}} + P_Y^{jX,vir}{q_{sX|Y}} + P_Z^{jX,vir}{q_{sX|Z}}} \right) + \lambda _{\min ,j}^{vir}} \right]\\
           &\le Y_{sX,jX}^{Z,vir} \le \frac{1}{4}{P_{j\alpha }}\left[ {{F_j}\left( {P_I^{jX,vir}{q_{sX|Id}} + P_X^{jX,vir}{q_{sX|X}} + P_Y^{jX,vir}{q_{sX|Y}} + P_Z^{jX,vir}{q_{sX|Z}}} \right){\rm{ + }}\lambda _{\max ,j}^{vir}} \right].
      \end{aligned}
  \end{equation}
  
  Only ${q_{sX|t}}$ is unknown in the above equation, and we still need to estimate its value in the practical process. Unlike the virtual process, the quantum state sent by Alice in the practical process is ${\left| {{\Phi _{j\beta }}} \right\rangle _{BE}}$. The yield of the practical states can be expressed as
  \begin{equation}
      Y_{sX,j\alpha }^{\alpha} = \frac{1}{{16}}{\mathop{\rm Tr}\nolimits} \left( {{{\hat D}_{sX}}\left| {{\Phi _{j\alpha }}} \right\rangle {{\left\langle {{\Phi _{j\alpha }}} \right|}_{BE}}} \right),
  \end{equation}
  where $\frac{1}{{16}}$ is that the joint probability that Alice sends one of the four states and Bob chooses the $X$ basis for his measurement. In the same way as the analysis of virtual process, we can obtain
  \begin{equation}
      \begin{aligned}
          &\frac{1}{{16}}\left[ {{M_{j\alpha }}\left( {{q_{sX|I}} + P_X^{j\alpha }{q_{sX|X}} + P_Y^{j\alpha }{q_{sX|Y}} + P_z^{j\alpha }{q_{sX|Z}}} \right) + {\lambda _{\min,j\alpha }}} \right]\\
          &\le Y_{sX,j\alpha }^{\alpha } \le 
          \frac{1}{{16}}\left[ {{M_{j\alpha }}\left( {{q_{sX|I}} + P_X^{j\alpha }{q_{sX|X}} + P_Y^{j\alpha }{q_{sX|Y}} + P_z^{j\alpha }{q_{sX|Z}}} \right) + {\lambda _{\max,j\alpha }}} \right],
          \end{aligned}
  \end{equation}
  where $\lambda_{\min,j\alpha}$ and $\lambda_{\max,j\alpha}$ are the two eigenvalues of the matrix $\left[ {\begin{array}{*{20}{c}}
      {{L_{j\alpha }}}&{{N_{j\alpha }}}\\
      {N_{j\alpha }^*}&{{P_{j\alpha }}}
      \end{array}} \right]$ (detailed in Appendix \ref{APPENDIX}).
  By substituting $j\alpha  \in \left\{ {0Z,1Z,0X,0Y} \right\}$ into the above equation, we can obtain a system of four linear inequalities, which is 
  \begin{equation}
      \begin{aligned}
          &{\left( {16\left[ {Y_{sX,0Z}^Z,Y_{sX,1Z}^Z,Y_{sX,0X}^X,Y_{sX,0Y}^Y} \right] - \left[ {\lambda _{\max,0Z },\lambda _{\max,1Z },\lambda _{\max,0X },\lambda _{\max,0Y }} \right]} \right)}\\
          &{ \le \left[ {{q_{sX|Id}},{q_{sX|X}},{q_{sX|Y}},{q_{sX|Z}}} \right]Q}\le\\
          &{\left( {16\left[ {Y_{sX,0Z}^Z,Y_{sX,1Z}^Z,Y_{sX,0X}^X,Y_{sX,0Y}^Y} \right] - \left[ {\lambda _{\min,0Z },\lambda _{\min,1Z },\lambda _{\min,0X },\lambda _{\min,0Y }} \right]} \right)},
      \end{aligned}
  \end{equation}
  where $Q: = {M_{j\alpha }}\left[ {{{\vec V}_{0Z}},{{\vec V}_{1Z}},{{\vec V}_{0X}},{{\vec V}_{0Y}}} \right]$ and ${{\vec V}_{j\alpha }}: = \left[ {1;P_X^{j\alpha };P_Y^{j\alpha };P_Z^{j\alpha }} \right]$.
  
\end{widetext}
 The range of the transmission rate ${q_{sX|t}}$ can be calculated by considering the practical process, and the upper (lower) bound on the denominator (numerator) in Eq. (\ref{Ephase}) can be found by substituting ${q_{sX|t}}$ into Eq. (\ref{virstate}). The phase error rate in $YY$, $XY$, $YX$ basis can also be calculated in the same way. As above, the lower bound of the $C$ value is found.

\begin{figure*}[htbp]
  \centering

  \subfloat[][setting $\delta_{IM}$ and $\delta_{PM}$ without non-qubit sources.]{
  \label{figa}
  \includegraphics[width=\mysize]{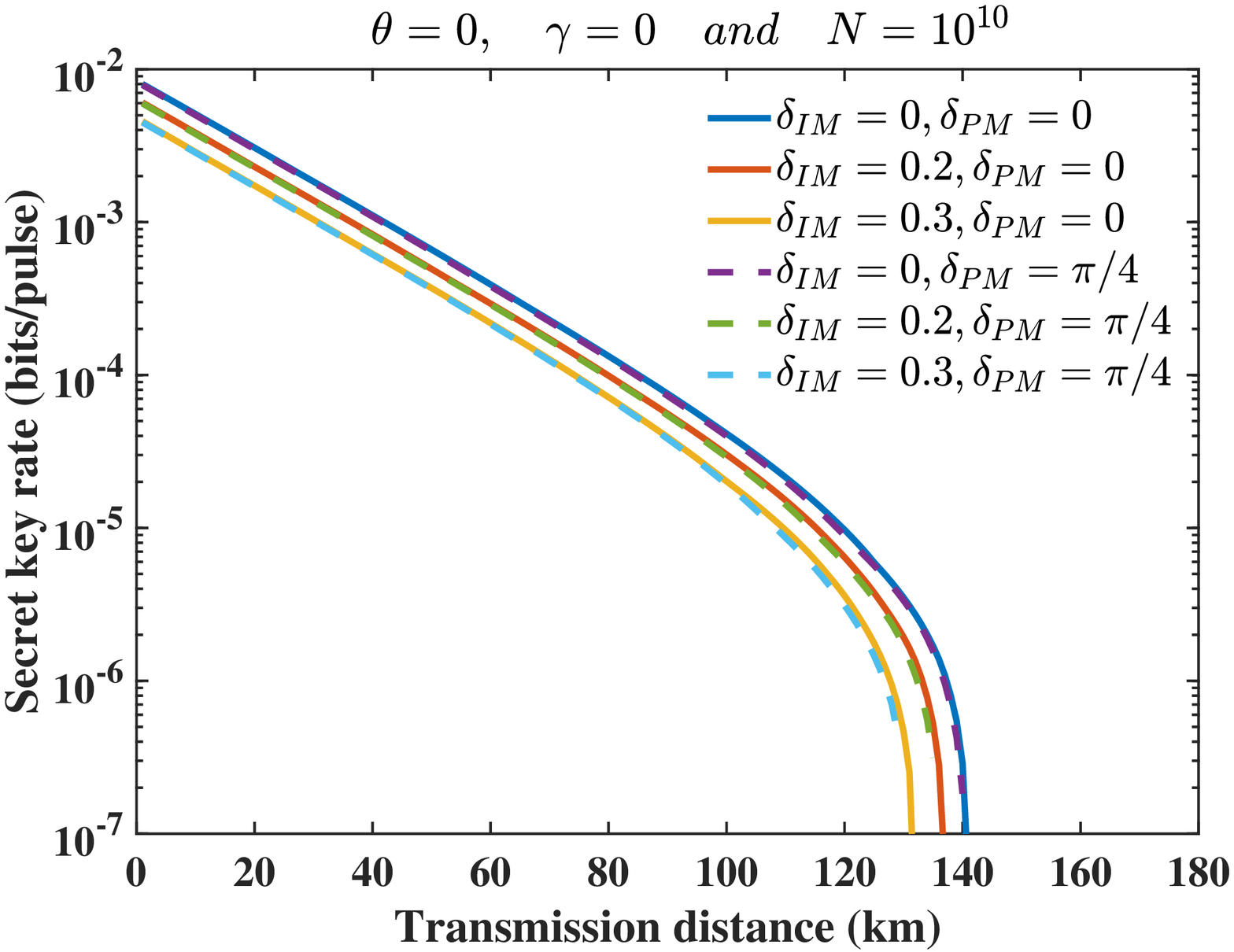}
  }
  \!
  \subfloat[][setting independent $\theta$ without the SPF.]{
  \label{figb}
  \includegraphics[width=\mysize]{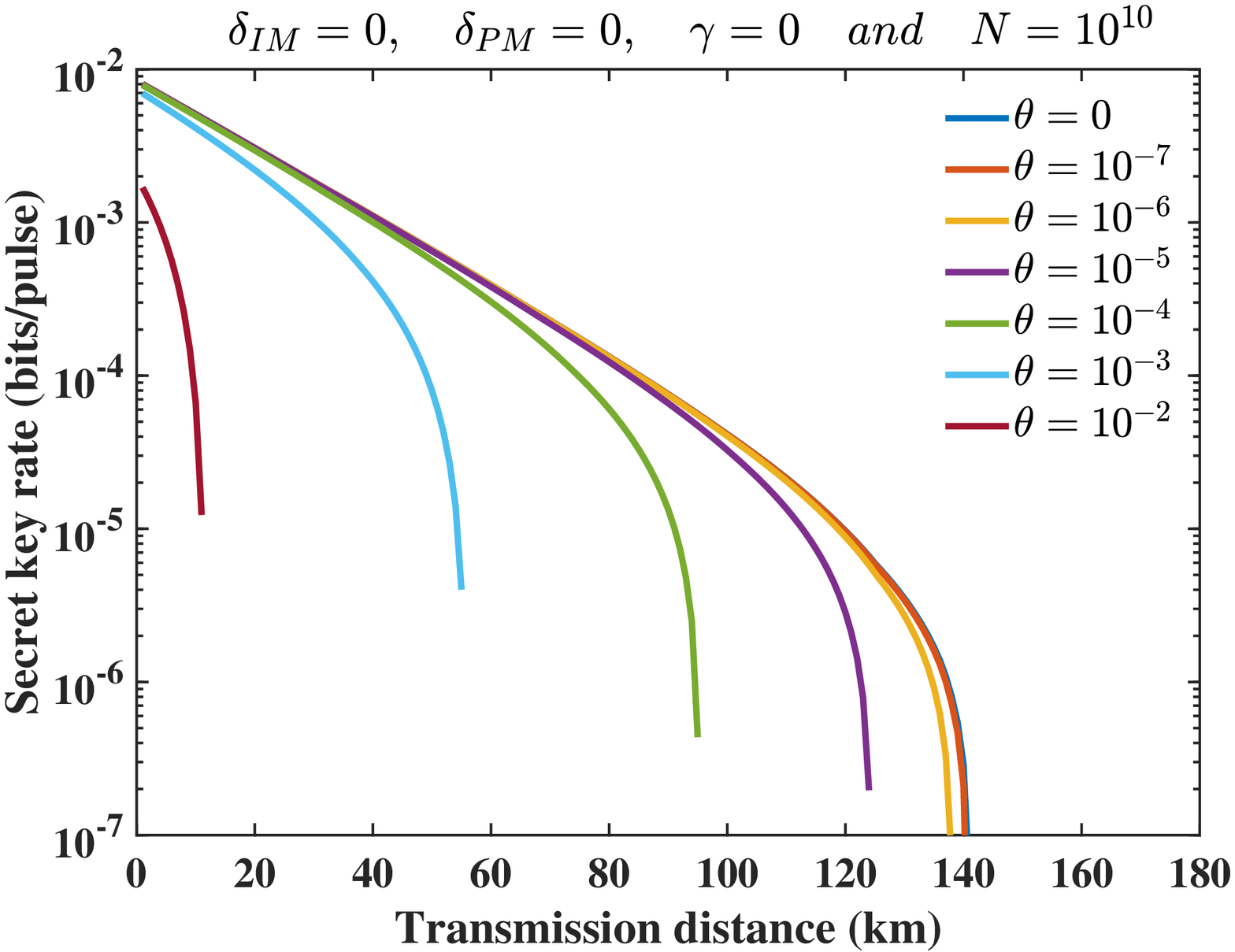}
  }
  \!
  \subfloat[][setting dependent $\hat \theta$ without the SPF;]{
  \label{figc}
  \includegraphics[width=\mysize]{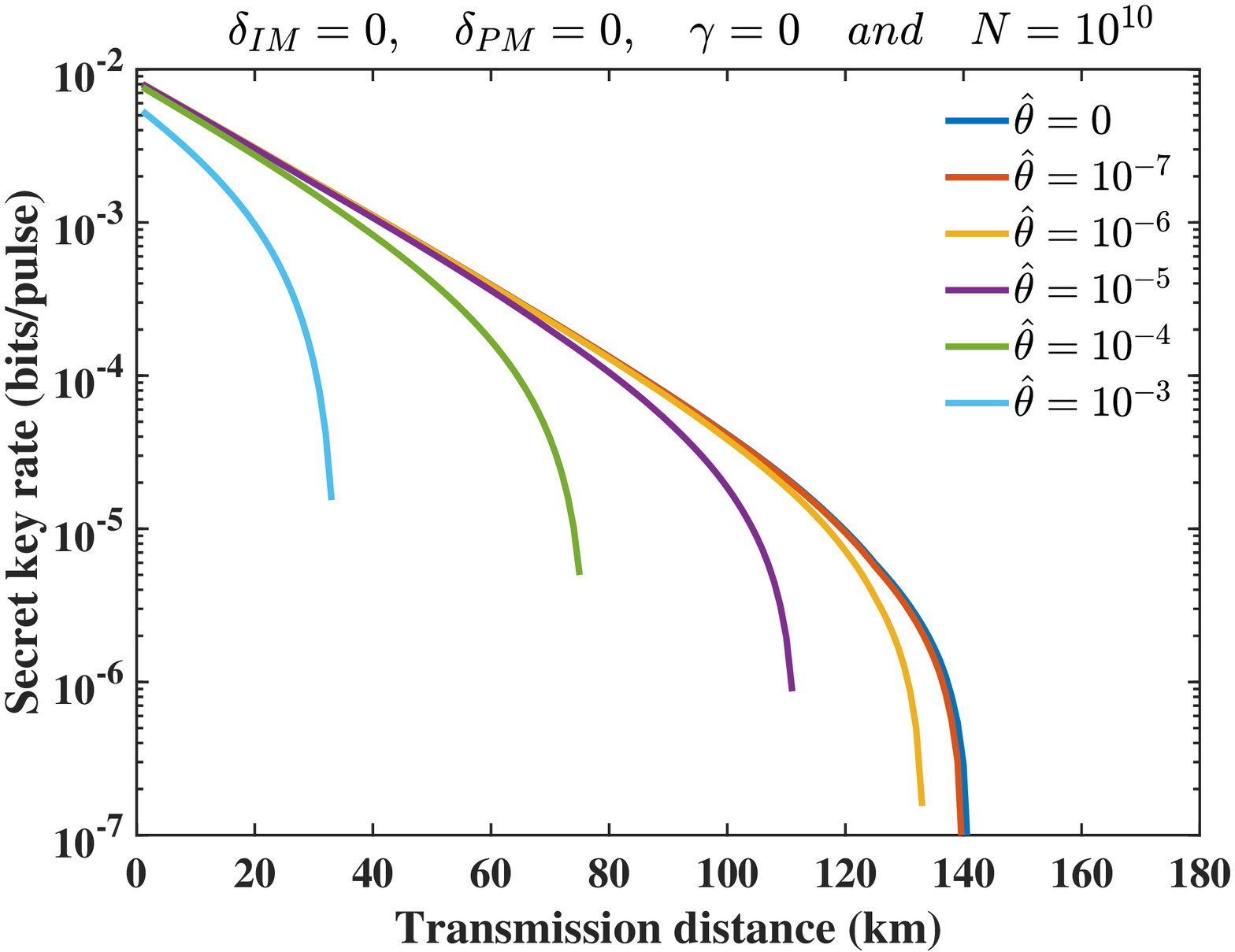}
  }
  \!
  \subfloat[][setting dependent $\hat \theta$ with the SPF.]{
  \label{figd}
  \includegraphics[width=\mysize]{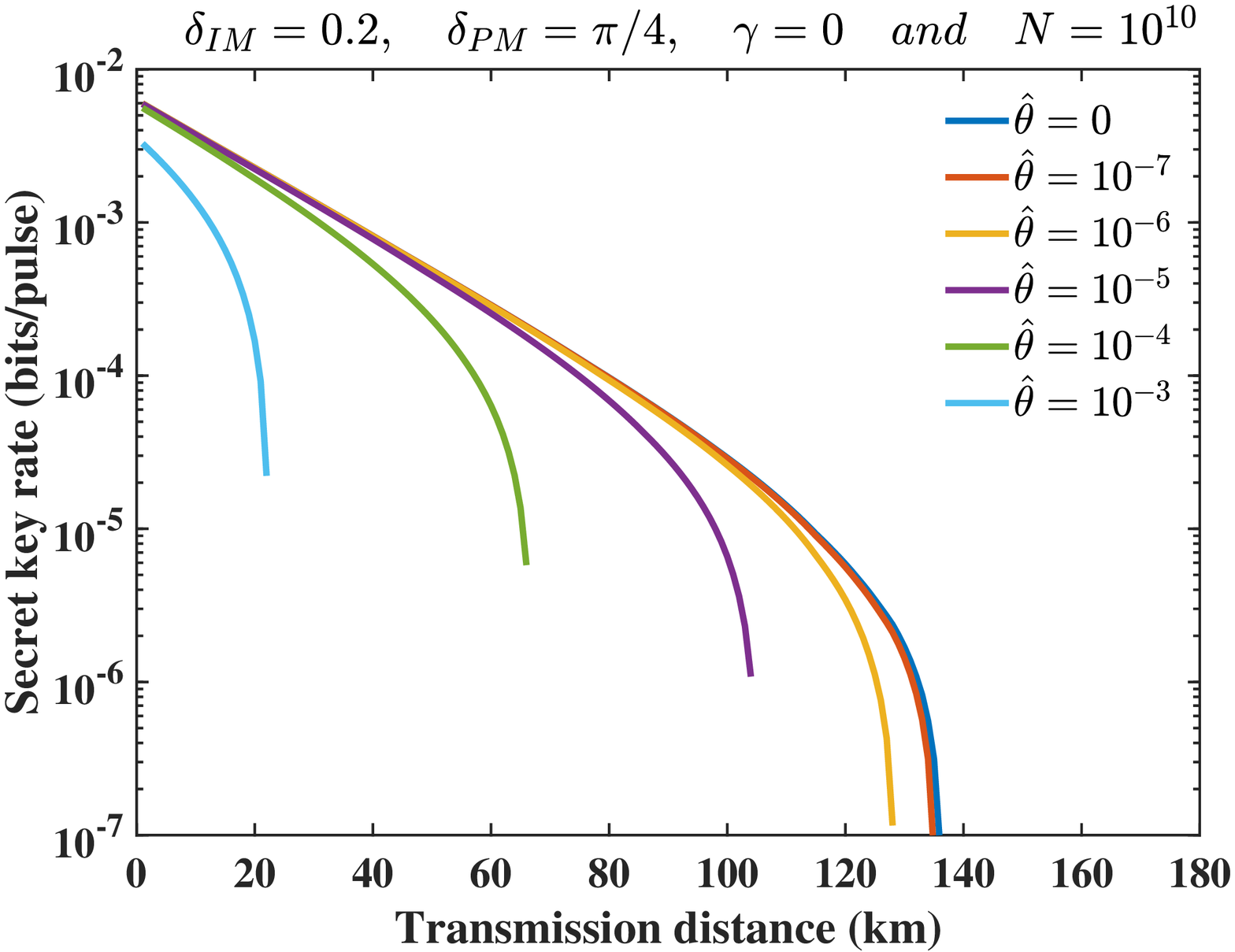}
  }
  \!
  \subfloat[][setting the reflected back Trojan photon $\gamma$ without the SPF.]{
  \label{fige}
  \includegraphics[width=\mysize]{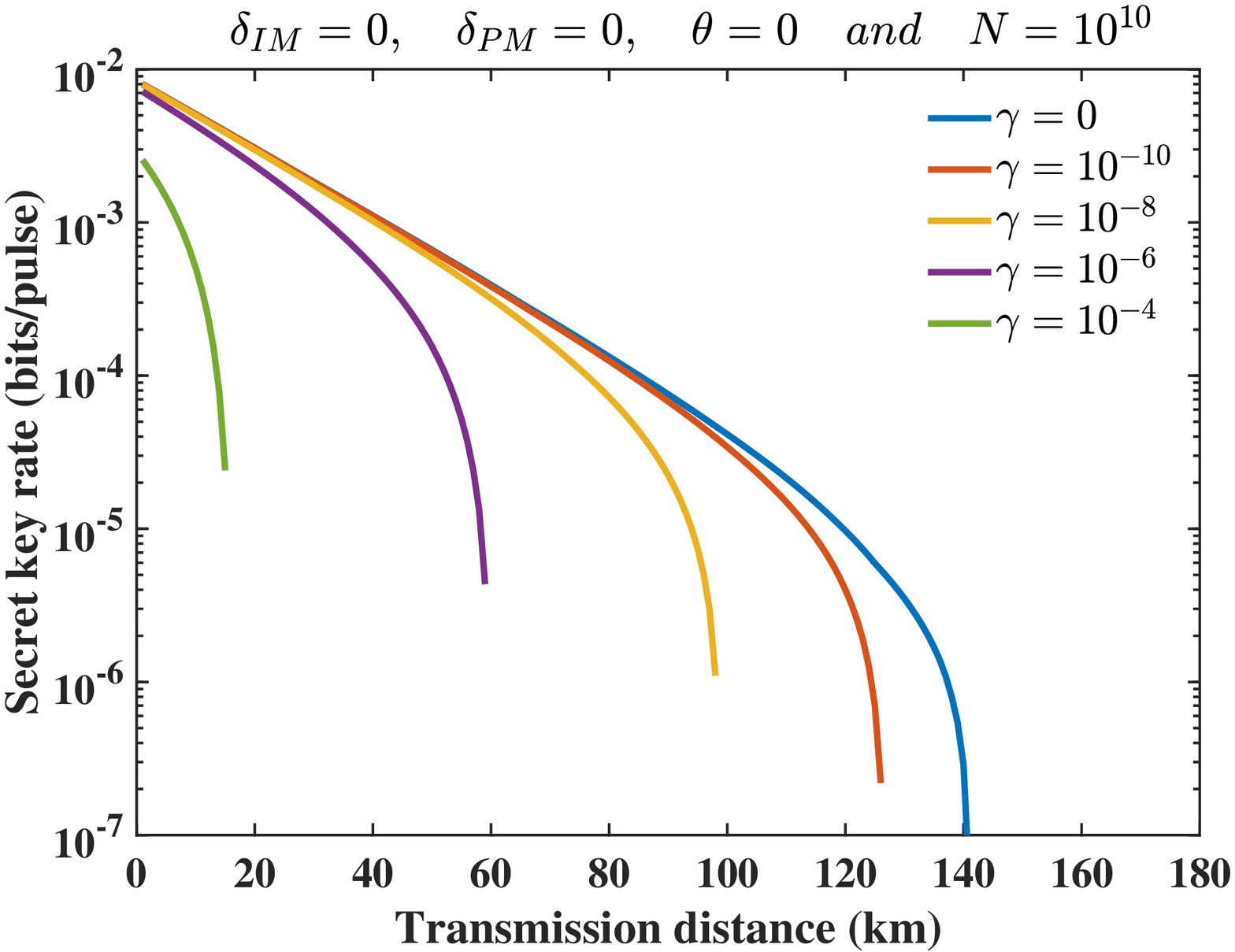}
  }
  \!
  \subfloat[][setting the total number of pulses $N$ with non-qubit sources.]{
  \label{figf}
  \includegraphics[width=\mysize]{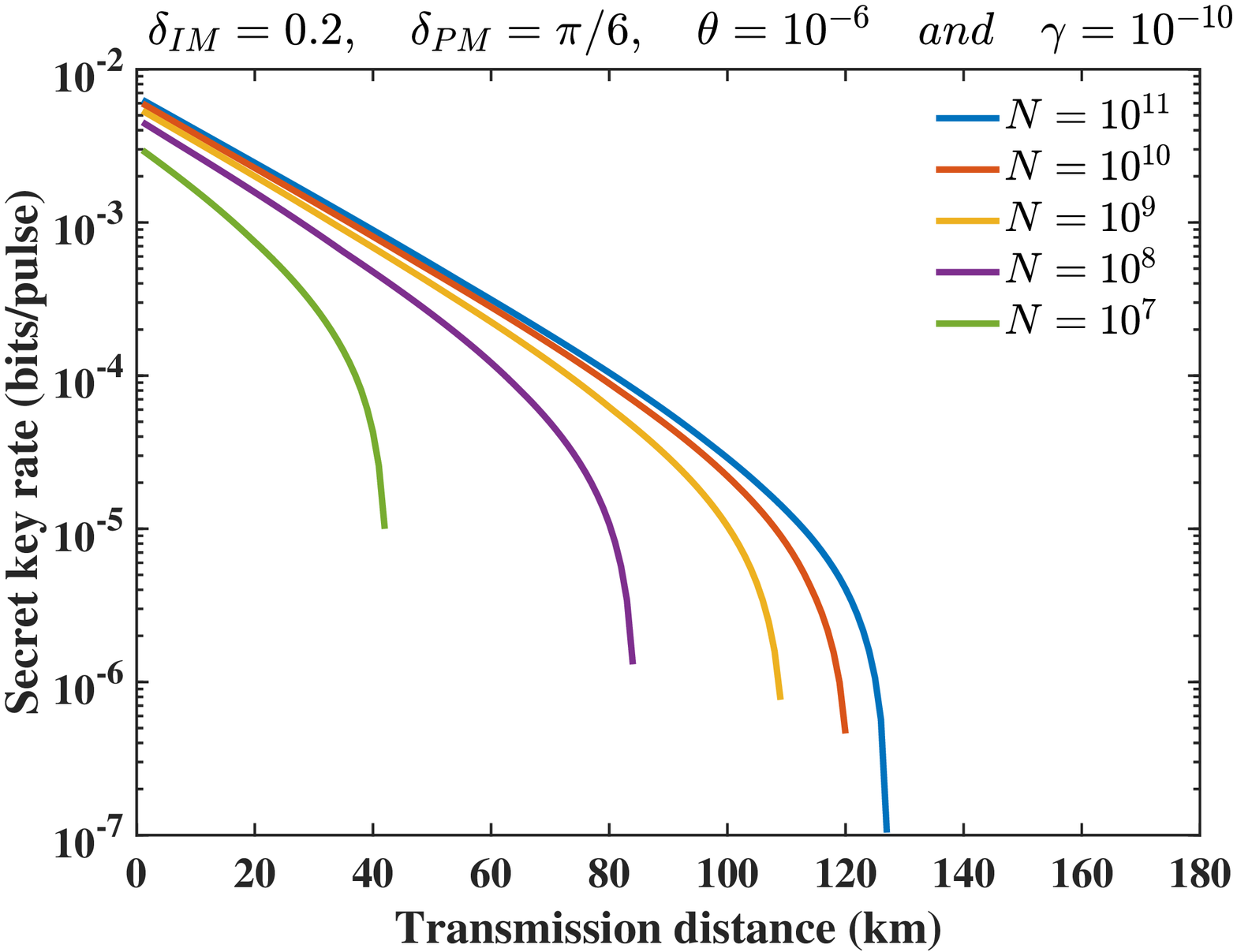}
  }
  \caption{
      \label{fig1}
      Secret key rates with finite-key analysis versus channel distance from Alice to Bob: we set different values
      of $\delta_{IM}$, $\delta_{PM}$, independent $\theta$, dependent $\hat \theta$, $\gamma$, and the total number of pulses $N$. }
  \end{figure*}

  \section{Simulation}

  In this section, we show the secret key rate with the SPF and non-qubit sources based on a phase-randomized WCS with a vacuum and weak decoy state in the finite-key analysis. The secret key rate in this case can be expressed as 
  \begin{equation}
      R = Q_{ZZ,1}^L\left( {1 - I_E^U} \right) - Q_{ZZ}^\mu {f_{EC}}h\left( {E_{ZZ}^{\mu ,bit}} \right).
  \end{equation}
  Here, $Q_{ZZ,1}^L$ is the lower bound of the single-photon
  counting rate when Alice and Bob choose $Z$ basis at the same time. Gain $Q_{ZZ}^\mu $ and bit error rate ${E_{ZZ}^{\mu ,bit}}$ are obtained when  Alice and Bob
  operate in the $Z$ basis and $\mu$ simultaneously. Moreover, ${f_{EC}}$ and $\mu$ are the error-correction coefficient and the intensity of the signal state.

  The channel model and finite-key analysis are described in Ref. \cite{PhysRevA.92.042319,PhysRevA.90.052314} and Ref. \cite{PhysRevA.89.022307,PhysRevA.99.032309}. We assume that the channel is not affected by the THA and the non-qubit assumption. Correlation coefficient settings including dark count $P_{d}$, fiber loss $\alpha$, detection efficiency $\eta_{D} $, finite-key coefficient $\varepsilon $ are described in detail in Table \ref{tab1}. The intensity of the weak coherent state is optimized to obtain the maximum secret key rate in Fig. \ref{fig1}.
       \begin{table}[htbp]
        \caption{
        Coefficients for Numerical Simulation
        }
        \begin{ruledtabular}
        \begin{tabular}{ccccc}
        $P_{d}$ & $\alpha$ & $\eta_{D} $ & $f_{EC}$ & $\varepsilon$\\
        \colrule
        $10^{-6}$ & $0.21$ & $0.2$ & $1.22$&$10^{-10}$ \\
        \end{tabular}
        \end{ruledtabular}
        \label{tab1}
        \end{table}   

   Figure. \ref{figa}  illustrates the impact of the SPF. For convenience, we assume that ${\delta _{IM1}} = {\delta _{IM2}} = {\delta _{BS1}} = {\delta _{BS1}} = {\delta _{IM}}$ and ${\delta _{PM1}} = {\delta _{PM2}} = {\delta _{PM}}$. Simulation results demonstrates that even if $\delta_{PM}$ is equal to the worst value ${\pi  \mathord{\left/
  {\vphantom {\pi  4}} \right.
  \kern-\nulldelimiterspace} 4}$, secret key rates are almost unaffected, which means that the system is not sensitive to the fluctuation of PM. With the increase of $\delta_{IM}$, the secret key rate and transmission distance decrease slightly, which indicates Eve cannot benefit from exploiting loss. Therefore, the SPF in IM and PM is tolerant, which is the advantage of the original LT protocol. 
  
  Figure \ref{figb} illustrates the case that $\theta_{j\alpha}$ and the choice of $j\alpha$ are independent, which implies that Eq. (\ref{polarization}) can be written as     ${\left| {{\Omega _{j\alpha }}} \right\rangle _{B}} = \cos {\theta }{\left| {{\phi _{j\alpha }}} \right\rangle _{HB}} + \sin {\theta }{\left| {{\phi _{j\alpha }}} \right\rangle _{VB}}$. The best case for secret key rates is when $\theta$ is equal to 0, because the single photon signal sent by Alice is a qubit. As the $\theta$ increases, the component of the vertical polarization direction $\sin \theta {\left| {{\phi _{j\alpha }}} \right\rangle _{VB}}$ also increases and Eve may have more opportunities to attack the state. At this moment, the encoding quantum state is in a three-dimensional Hilbert space, and Eve can perform unambiguous state discrimination attacks on it. The simulation results illustrate that the effect of $\theta$ on the secret key rate can be negligible as long as $\theta$ is less than $10^{-6}$. If Alice can recognize the value of $\theta$ well, Eve cannot use this vulnerability to eavesdrop any information. 
  
  Figure \ref{figc} illustrates the case that $\theta_{j\alpha}$ and the choice of $j\alpha$ are dependent. The $Z$ basis is encoded with IM, while $X$ and $Y$ basis are encoded with PM in the time-bin phase coding system. It is noted that IM can be identified as a Mach-Zehnder interferometer with encoded $\pi$ phase when encoding $Z$ basis. Thus, we can simply assume that
  \begin{equation}\label{xiangguan}
      \begin{aligned}
          {\theta _{0Z}} &= \left( {\frac{{{\delta _{IM1}}}}{2} + \pi } \right)\hat \theta ,&&{\theta _{1Z}} = \left( {\frac{{{\delta _{IM2}}}}{2} + \pi } \right)\hat \theta ,\\
          {\theta _{0X}} &= {\delta _{PM1}}\hat \theta ,&&{\theta _{0Y}} = \left( {{\delta _{PM2}} + \frac{\pi }{2}} \right)\hat \theta .
      \end{aligned}
  \end{equation}
  As shown in Fig. \ref{figc}, the secret key rate is reduced compared to the independent $\theta$. Considering the SPF, the decline of the secret key rate is more obvious in Fig. \ref{figd}.
  
  The effect of different reflected back photon intensity $\gamma$ in THA are indicated in Fig. \ref{fige}. Obviously, the secret key rate decreases significantly with the increase of $\gamma$, and Eve can extract more amount of information from the reflected back Trojan photon. Alice can be equipped with isolation equipment so that $\gamma <10^{-10}$, which effectively blocks THA attacks.
  
  Finally, in practice, a finite number of quantum states are distributed. The results for different finite-key lengths are shown in Fig. \ref{figf}.
  
  \section{conclusion}
  In conclusion, we have analyzed the RFI-QKD protocol based on generalizing loss-tolerant technique. Compared to the original RFI-QKD protocol, our work removes  unrealistic assumptions, in other words, it is able to operate with non-qubit sources. In the GLT-RFI-QKD protocol, the phase error rate with each basis and the information eavesdropped by Eve are estimated by a virtual process. The simulation results show that multiple optical modes caused by Trojan horse attacks and equipment loopholes cannot be ignored. As long as prepared quantum states (including $\delta_{IM}$, $\delta_{PM}$, independent $\theta$, dependent $\hat \theta$, and $\gamma$) can be appropriately characterized, we can obtain high-quality secure keys in the GLT-RFI-QKD protocol. Moreover, only four quantum states need to be prepared instead of six, which greatly reduces the complexity of the experiment. 
  
  The method can also be used in other protocols such as quantum secure direct communication(QSDC) \cite{PhysRevA.65.032302}, because the essential idea is to put the mis-alignment in the frame of reference into the channel capacity \cite{https://doi.org/10.1002/que2.26}, which is equivalent to secure key rate in QKD. It is also applicable to MDI protocols such as MDI-QKD, twin-field QKD \cite{YIN202193}, and MDI-QSDC \cite{ZHOU202012,RN22}. 
  
\begin{acknowledgments}
  This work has been supported by Fundamental Research Funds for the Central Universities No. 2019XD-A02; State Key Laboratory of Information Photonics and Optical Communications (Beijing University of Posts and Telecommunications) No. IPOC2019ZT06; BUPT innovation and entrepreneurship support program No. 2021-YC-A315.
\end{acknowledgments}

\appendix
\section{Explanation of coefficients}
\label{APPENDIX}  
    In this section, we provide detailed expressions for some of the coefficients in \ref{Calculation of parameter C}. Coefficients in the practical process are as follows:
\begin{equation}
    \begin{aligned}
        {M_{j\alpha }} &= {\cos ^2}{\theta _{j\alpha }}T_I^2,{L_{j\alpha }} = {\sin ^2}{\theta _{j\alpha }}\left( {T_I^2 + T_D^2} \right),\\
        {N_{j\alpha }} &= \cos {\theta _{j\alpha }}\sin {\theta _{j\alpha }}T_I^2,{P_{j\alpha }} = {\cos ^2}{\theta _{j\alpha }}T_D^2,\\
        {{\vec V}_{0Z}} &= \left[ {1;\sin {\delta _{IM1}};0;{\rm{cos}}{\delta _{IM1}}} \right],\\
        {{\vec V}_{1Z}} &= \left[ {1;\sin {\delta _{IM2}};0; - {\rm{cos}}{\delta _{IM2}}} \right],\\
        {{\vec V}_{0X}} &= \left[ {1;\cos {\delta _{BS1}}\cos {\delta _{PM1}};\cos {\delta _{BS1}}\sin {\delta _{PM1}};\sin {\delta _{BS1}}} \right],\\
        {{\vec V}_{0Y}} &= \left[ {1; - \cos {\delta _{BS2}}\sin {\delta _{PM2}};\cos {\delta _{BS2}}\cos {\delta _{PM2}};\sin {\delta _{BS2}}} \right].
    \end{aligned}
\end{equation}
Coefficients in the virtual process are as follows:
\begin{widetext}
    \begin{equation}
        \begin{aligned}
                {F_j} =& T_I^2\left( {\cos \theta _{0Z}^2{\rm{ + }}{{\left( { - 1} \right)}^j}\sin \left( {\frac{{{\delta _{IM1}}}}{2} + \frac{{{\delta _{IM2}}}}{2}} \right)\cos {\theta _{0Z}}\cos {\theta _{1Z}} + \cos \theta _{1Z}^2} \right),\\
                {H_j} =& T_I^2\left( {{{\sin }^2}{\theta _{0Z}}{\rm{ + }}{{\left( { - 1} \right)}^j}\sin \left( {\frac{{{\delta _{IM1}}}}{2} + \frac{{{\delta _{IM2}}}}{2}} \right)\sin {\theta _{0Z}}\sin {\theta _{1Z}}+{{\sin }^2}{\theta _{1Z}}} \right) + 2T_D^2,\\
                {G_j} =& \sqrt {{F_j}{H_j}}.
        \end{aligned}
    \end{equation}  
\end{widetext}

\bibliography{ref}

\end{document}